\documentstyle[12pt,epsf,epsfig,wrapfig]{article}
\begin{document}
\begin{center}
{\large \bf
Equivalence of Covariant and Light-Front Perturbation Theory
\\ }
\vspace{5mm}
\underline{N.C.J. Schoonderwoerd} and
B.L.G. Bakker
\\
\vspace{5mm}
{\small\it
Vrije Universiteit, De Boelelaan 1081, 1081 HV Amsterdam, The Netherlands
\\ }
\end{center}

\begin{center}
ABSTRACT

\vspace{5mm}
\begin{minipage}{130 mm}
\small
Light-Front Field Theory (LFFT) is a good candidate to describe bound states.
In LFFT covariance is non-manifest. Burkardt and Langnau[1] claim that, even
for scattering amplitudes, rotational invariance is broken.
We will take a different path of obtaining rules for light-front time-ordered
diagrams[2]. Covariance depends on the choice of a regulator
$\alpha$. We need to apply some regularisation scheme to render 
physical amplitudes finite. This is done by applying minus regularisation[3]. In
this process all ambiguities related to the regulator $\alpha$ are removed. 
Therefore there is equivalence between the perturbative expansions
of covariant and LFFT.  
\end{minipage}
\end{center}

\section{Introduction}

Covariant Field Theory (CFT) has been very successful in describing
elementary particles but has not produced a convenient framework
to describe bound states of elementary particles. However,
Hamiltonian field theories seem to be good candidates to
explain properties of bound states. In a Hamiltonian frame work
the initial conditions are specified on some surface.
The Hamiltonian then gives the evolution of the system in time. 
Already in 1949, Dirac pointed out that there are several
possible choices for the surface of quantisation. One of these surfaces
is the light-front. There are a number of advantages to LFFT over
quantisation on, e.g., the equal-time plane. 
In LFFT there can be no creation/annihilation of massive particles 
from/to the
vacuum. This reduces the number of time-ordered diagrams. 
However, we have to add so-called instantaneous terms 
for every fermion line, because we use on-shell spinors.
For a number of reasons, quantisation on the light-front is
very complicated. In Naive Light-Cone Quantisation (NLCQ)
some problems are not satisfactorily solved.
Still, NLCQ rules have been constructed for 
light-front time (lime) ordered diagrams. Inspired by Ligterink[2] 
we will construct rules for lime-ordered diagrams
in another way. 
Ligterink derived rules for lime-ordered diagrams by integrating
covariant Feynman diagrams over light-front time. 
For some types of diagrams the integrals diverge.
So, only upon regularisation these integrals, the relation becomes
definite.

In LFFT, or any other Hamiltonian theory, covariance 
is not manifest. Burkardt and Langnau[1] claim that in NLCQ
rotational invariance is broken.
They fix rotational invariance by introducing non-covariant
counterterms. 
Transverse divergences are dealt with using dimensional regularisation. 
Instead, we will use the method of minus regularisation[3]. 
In this method longitudinal and transverse divergences are treated 
in the same way. 
Another advantage of minus regularisation is that the ambiguity, caused by 
integration over light-front time, is removed. This means that LFFT
is equivalent to CFT in perturbation theory and that our method will yield 
covariant physical amplitudes.

\section{Calculation of the fermion self energy}
As an example we will show equivalence for one diagram in the Yukawa model.
Our light-front coordinates are defined as: $k^\pm = (k^0 \pm k^3)/\sqrt{2}$.
The self interaction of a fermion with mass $m$
and momentum $q$
via a boson with mass $l$ is given by
\label{sresidue}
\begin{equation}
\hspace{-.3cm}
{\epsfxsize=3cm \epsffile[-10 10 150 70]{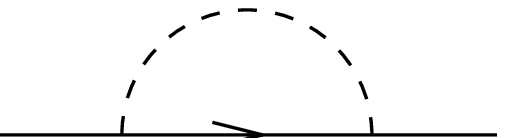} }
\hspace{-.3cm}= - \hspace{-.1cm}\int 
\frac{{\rm d}^{2}k^\perp {\rm d}k^+ {\rm d}k^-}{4 k^+ (q^+ - k^+)}
\frac{ k^- \gamma^+ + k^+ \gamma^- - k^\perp \gamma^\perp + m}
{(k^- \!- \frac{{k^\perp}^2 + m^2 - i \epsilon}{2 k^+}) 
 (q^- \!\!-\! k^- \!-\! \frac{(q^\perp - k^\perp)^2 + l^2 - i \epsilon}{2 (q^+ - k^+)})}
\end{equation}
The integral (1) is not defined and we insert the following
regulator to eliminate the pole at infinity. 
\begin{equation}
\frac{\alpha(k^+)}{1 + i \delta q^+ k^-}
+ \frac{ 1 - \alpha(k^+)}{1 - i \delta q^+ k^-}
\end{equation}
It is convenient to take $\alpha(k^+) = 0$ for $k^+ < 0$ and 
$\alpha(k^+) = 1$ for $k^+ > q^+$
to simplify the contour integration. 
After taking the limit $\delta \rightarrow 0$
we find
\begin{equation}
{\epsfxsize=3cm \epsffile[-10 10 150 70]{fse.eps} }
=
{\epsfxsize=3cm \epsffile[-10 20 150 80]{fseprop.eps} }
+
{\epsfxsize=3cm \epsffile[-10 20 150 80]{fseinst.eps} }
\end{equation}
with the propagating part 
\begin{equation}
\hspace{-.3cm}
\epsfxsize=3cm \epsffile[-10 20 150 80]{fseprop.eps} 
\hspace{-.3cm}=
2 \pi i 
\int {\rm d}^{2}k^\perp
\int_0^{q^+} \frac{{\rm d}k^+}{4 k^+ (q^+ - k^+)}
\frac{ \frac{m^2 + {k^\perp}^2}{2 k^+}
 \gamma^+ + k^+ \gamma^- - k^\perp \gamma^\perp + m}
{q^- - \frac{m^2 + {k^\perp}^2}{2 k^+} -
      \frac{l^2 + (q^\perp - k^\perp)^2}{2(q^+ - k^+)}}
\end{equation}
and the instantaneous part
\begin{equation}
\epsfxsize=3cm \epsffile[-10 20 150 80]{fseinst.eps}
= 2 \pi i 
\int {\rm d}^{2}k^\perp
\int_0^{q^+} {\rm d}k^+
(1 - \alpha(k^+)) \frac{\gamma^+}{4 k^+ (q^+ - k^+)}
\end{equation}
We see that (4) has indeed the usual form for a propagating lime-ordered
diagram with an on-shell spinor in the numerator. 
In (5) we see the instantaneous part containing an
extra factor depending on the regulator. At this level equivalence
between covariant and LFFT is ambiguous because it depends
on the regulator $\alpha$. We will see that this dependence is
removed naturally upon using minus regularisation.

\section{Minus regularisation}

The propagating and the instantaneous contributions suffer from
both longitudinal and transverse divergences. If these divergences
are treated in different ways we should not be surprised that it is hard
to recover covariance. However, we will use the minus regularisation
scheme[3] which treats these divergences on the same footing. 
It removes the lowest orders in the Taylor expansion of the amplitude.
We differentiate the diagram with respect to the external energy
until the integration is finite. After integrating over the internal
momenta  
we integrate the result as many times with respect to the external energy
as we have differentiated
before. 
The propagating part of the fermion self energy (4) contains a term 
proportional to ${k^\perp}^2/2k^+$ which has to be differentiated twice.
The operation we perform is then 
\begin{equation}
\int_{\frac{q_\perp^2}{2 q^+}}^{q^-} {\rm d}q' 
\int_{\frac{q_\perp^2}{2 q^+}}^{q'^-} {\rm d}q''
\int {\rm d}^{2}k^\perp
\int_0^{q^+} {\rm d}k^+
\left( \frac{\partial}{\partial q''^-} \right)^2
\end{equation}
The other terms in the numerator of (4) must only be differentiated once.  
Otherwise the subtracted terms would not be local and would 
not correspond to counterterms in the Lagrangian. Apparently, we need 
to discriminate between different parts of the same diagram. 
The instantaneous diagram (5) has to be differentiated once to remove the
singularity. Since the integrand is independent of the energy $q^-$ the
differentiation kills the integrand. 
Therefore the $\alpha$ dependence is lost.

\section{Conclusions}

At the level of the unregularised diagrams equivalence is obscured by
longitudinal divergences.  These divergences are dealt with using a
regulator $\alpha$. Upon using minus regularisation the $\alpha$
dependence, and therefore the ambiguity, is removed.  Since the minus
regularisation is a linear operation it commutes with the expansion of
the  covariant diagram in lime-ordered diagrams. Therefore, the
regularised perturbation series are equivalent.  Our procedure can be
generalised to any diagram in the Yukawa model.  Minus regularisation
removes local terms in the series of lime-ordered diagrams. Those local
terms are the diagrams containing the regulator. 
So, equivalence is restored and therefore LFFT will yield covariant
physical amplitudes.

\vspace{0.2cm}
\vfill
{\small\begin{description}
\item{[1]} 
M. Burkardt and A. Langnau,
Rotational invariance in light-cone quantization,
Physical Review {\bf D44} (1991) 3857.
\item{[2]}
N. E. Ligterink and B. L. G. Bakker,
Equivalence of light-front and covariant field theory,
Physical Review {\bf D52} (1995) 5954.
\item{[3]}
N. E. Ligterink and B. L. G. Bakker,
Renormalization of light-front Hamiltonian field theory,
Physical Review {\bf D52} (1995) 5917.
\end{description}}

\end{document}